\documentclass[conference]{IEEEtran}
\IEEEoverridecommandlockouts
\usepackage{cite}
\usepackage{amsmath,amssymb,amsfonts}
\usepackage{algorithmic}
\usepackage{graphicx}
\usepackage{textcomp}
\usepackage{mathrsfs}
\usepackage{xcolor}

\def\BibTeX{{\rm B\kern-.05em{\sc i\kern-.025em b}\kern-.08em
    T\kern-.1667em\lower.7ex\hbox{E}\kern-.125emX}}
\def\endthebibliography{%
  \def\@noitemerr{\@latex@warning{Empty `thebibliography' environment}}%
  \endlist
}
\begin{document}
\title{MARL for Decentralized Electric Vehicle Charging Coordination with V2V Energy Exchange
\thanks{This work was supported in part by the Australian Research Council (ARC) Discovery Early Career Researcher Award (DECRA) under Grant DE230100046.}
}
\author{\IEEEauthorblockN{Jiarong Fan$^{1,2}$, Hao Wang$^{1,2*}$, Ariel Liebman$^{1,2}$}
\IEEEauthorblockA{
    $^{1}$\textit{Department of Data Science and AI, Faculty of IT, Monash University, Melbourne, VIC 3800, Australia} \\
    $^{2}$\textit{Monash Energy Institute, Monash University, Melbourne, VIC 3800, Australia} \\
    Emails: \{jiarong.fan, hao.wang2, ariel.liebman\}@monash.edu}
\thanks{*Corresponding author: Hao Wang.}
}

\maketitle

\begin{abstract}
Effective energy management of electric vehicle (EV) charging stations is critical to supporting the transport sector's sustainable energy transition. This paper addresses the EV charging coordination by considering vehicle-to-vehicle (V2V) energy exchange as the flexibility to harness in EV charging stations. Moreover, this paper takes into account EV user experiences, such as charging satisfaction and fairness. We propose a Multi-Agent Reinforcement Learning (MARL) approach to coordinate EV charging with V2V energy exchange while considering uncertainties in the EV arrival time, energy price, and solar energy generation. The exploration capability of MARL is enhanced by introducing parameter noise into MARL's neural network models. Experimental results demonstrate the superior performance and scalability of our proposed method compared to traditional optimization baselines. The decentralized execution of the algorithm enables it to effectively deal with partial system faults in the charging station.
\end{abstract}


\section{Introduction}
Electric vehicles (EVs) have emerged as an effective solution to the net-zero transition of the transport sector, contributing to emission reduction and climate change mitigation. EV charging predominantly from renewable energy sources (RES) can further decarbonize the sector. Recent advances in charging technologies, e.g., vehicle-to-vehicle (V2V), have made EV charging more flexible, such that the use of intermittent RES can be increased. But integrating RES and V2V flexibility into the EV charging coordination process presents significant challenges. More specifically, the intermittent nature of RES and other exogenous uncertainties, such as EV charging behaviors and electricity prices, make EV charging optimization a challenging task. Moreover, EV coordination usually relies on communications between EVs to exchange operating information for making proper scheduling decisions. But given the time-varying environment, EV coordination through extensive communications is often impractical. Therefore, this paper is motivated to develop an effective EV charging coordination algorithm that can handle various uncertainties in the system without the need of extensive communications and information exchange between EVs.

There has been a large body of literature on EV charging coordination and energy management under uncertainties. In terms of the underlying methodology, existing studies can be classified into model-based approaches \cite{shurrab2021efficient, koufakis2019offline, razi2022smart} and model-free approaches \cite{cao2021smart, dorokhova2021deep, ye2022learning, qiu2022hybrid, Cooperative}. The model-based approaches rely on the precise modeling of the system and uncertainties. For example, \cite{koufakis2019offline} and \cite{razi2022smart} presented a centralized model-based scheduling method for EV charging coordination with V2V, and used the prediction of real-time energy prices. These methods heavily depend on the accuracy of predictions and could suffer from significant performance degradation when the prediction is less effective. Model-free approaches, mainly using deep reinforcement learning (DRL), have demonstrated substantial potential in addressing sequential decision-making problems under uncertainties. This provides a promising solution to handling uncertainties in EV charging coordination. For example, a customized actor-critic reinforcement learning algorithm was proposed in \cite{cao2021smart} to coordinate EV charging under uncertainties in EV charging behaviors.

DRL-based EV charging management is effective in handling uncertainties, but existing centralized solution methods rely on excessive communication between EVs, requiring reliable EV charging infrastructure and thus creating possible barriers to their implementation in practice. This concern is especially pertinent given the frequent occurrence of partial system faults in modern EV charging stations \cite{rempel2022reliability}. Hence, there is a compelling need for a new paradigm for DRL-based EV charging coordination that can obviate the necessity for excessive information exchange, consequently bolstering the system reliability. Recent research effort has been made to address this issue by introducing decentralized algorithms for EV charging coordination. For example, \cite{da2019coordination,Cooperative,Hybrid, ye2022learning} employed multi-agent reinforcement learning (MARL) to enable decentralized execution of EV charging decisions. But many decentralized EV charging methods still required some information exchange.

More advanced decentralized MARL algorithms have been proposed, e.g., in \cite{Hybrid}, to manage EV charging without information exchange. However, the results showed that many EV charging requests were not completed when EVs left the charging station. This problem is largely caused by the reward signal in MARL, and decentralized EV coordination could make this problem more challenging. In addition, uncompleted EV charging requests can cause concerns to EV owners, in particular, if there is a biased pattern in completing charging requests causing fairness issues.
According to \cite{danner2021quality}, EV user satisfaction with the charging process and the equitable provision of charging services can significantly affect the adoption of EVs. As such, it is imperative to develop decentralized EV coordination algorithms that are capable of navigating these intricate challenges.

This paper is motivated to bridge the aforementioned research gaps by developing an effective decentralized MARL algorithm for coordinating EV charging with V2V. A group of EVs arrive and depart with random charging demands, and the system manages EV charging using local RES and grid power with possible assistance from V2V which enables energy exchange among EVs when needed. The system also incorporates EV user satisfaction by considering charging completion metrics while ensuing fairness among EVs. The key contributions of this paper are as follows.
\begin{itemize}
    \item \textit{Decentralized EV coordination via MARL:} We develop a decentralized MARL algorithm to minimize energy costs of EVs under uncertainties in solar power generation, energy prices, and EV charging behaviors. Our algorithm is proficient in coordinating EV charging under uncertainties and enhances system reliability under system faults, such as partial EV charger failures.
    \item \textit{Fairness-aware user satisfaction:} We introduce a fairness model for charging satisfaction among EV users, which is subsequently incorporated into the design of DRL rewards. The results demonstrate the efficacy of this fairness model in optimizing the EV charging process.
    \item \textit{Noisy network for better exploration in MARL training:} We employ a noisy network as opposed to action noise, which fosters agent exploration and thereby accelerates convergence. Numerical results show that the proposed framework improves convergence during MARL training.
\end{itemize}

\section{Problem Formulation}\label{sec:problem}
We consider an EV charging station, which serves a set of EVs $\mathcal{I} {=} {1,..., I }$ connected to their corresponding chargers over a predetermined operational horizon $\mathcal{T}$. The objective of the system is to minimize the aggregate energy cost of the charging station.

\subsubsection{EV Charging Model}
We assume that EV chargers support bidirectional energy flow, and we split the decision variable of EVs into charging power and discharging power. We let $p_{i,t}^{\text{ch}}$ and $p_{i,t}^{\text{disch}}$ denote charging and discharging power for EV $i \in \mathcal{I}$ at time $t$. The constraints associated with these decision variables are as follows
    \begin{align}
        &0 \leq p_{i,t}^\text{ch} \leq (1-z_{i,t})\bar{P}^\text{ch}\label{xt_con},\\
        &0 \leq p_{i,t}^\text{disch} \leq z_{i,t}\bar{P}^\text{disch} \label{xdt_con},\\
        & z_{i,t} \in \{0,1\} \label{z_it},
    \end{align}
where $\bar{P}^\text{ch}$ and $\bar{P}^\text{disch}$ are the maximum charging and discharging power of EV chargers. We use binary variable $z_{i,t}$ to ensure that charge and discharge can not happen simultaneously.

During the EV charging process, the battery dynamic must be defined for setting battery constraints. The battery energy level $E_{i,t}$ and battery constraints for the $i$-th EV at $t$ can be expressed as
\begin{align}    
        &E_{i,t} = E_{i,t-1}+ p_{i,t}^\text{ch} \eta^{\text{ch}}_i \Delta t - \frac{p_{i,t}^\text{disch}\Delta t} {\eta^{\text{disch}}_i} \label{eit},\\
        &A_{i,t} E_i^\text{min}\leq E_{i,t} \leq A_{i,t} E_i^\text{max}, \label{eit_con}
\end{align}
where $\eta^{\text{ch}}_i$ and $\eta^{\text{disch}}_i$ denote the charging and discharging efficiencies of EV $i$, respectively. In \eqref{eit_con}, $E_i^\text{min}$ and $E_i^\text{max}$ are the minimum/maximum energy level of the $i$-th EV. The battery dynamic at one time step $\Delta t$ are represented in \eqref{eit} and $A_{i,t}$ indicates the status of the $i$-th EV at $t$. If $A_{i,t}$ is 0, the charger disconnects to an EV, which means the battery energy level $E_{i,t}$ is always 0. In order to satisfy the constraint (\ref{eit_con}), $p_{i,t}^\text{ch}$ and $p_{i,t}^\text{disch}$ must equal to 0, when $A_{i,t}$ is 0.


It is crucial to consider the additional battery degradation costs associated with EV discharging, as it affects the economic feasibility of EV charging/discharging \cite{UDDIN2018342}. According to \cite{wu2022optimal}, the energy throughput equivalent method, which primarily determines the Equivalent Full Cycle (EFC) for calculating the $i$-th EV's cycle aging $AGE_{i,t}^\text{cyc}$, is shown below
\begin{align}
    &EFC_{i,t} = 0.5 \cdot \frac{\lvert p_{i,t}^\text{ch} \eta^{\text{ch}}_i \Delta t -  \frac{p_{i,t}^\text{disch}\Delta t} {\eta^{\text{disch}}_i} \rvert}{E_i^\text{cap}}, \label{EFC}\\
    &AGE_{i,t}^\text{cyc} = \frac{EFC_{i,t}}{L_i^\text{cyc}}, \label{AGE}
\end{align}
in which $E_i^\text{cap}$ represents the battery capacity of the $i$-th EV and 0.5 signifies the mean half cycle in the battery life. The proportion of battery aging in a time step in the total battery cycle life, denoted as $L_i^\text{cyc}$, can be computed by Equation \eqref{AGE}. This signifies the cycle aging, $AGE_{i,t}^\text{cyc}$, for the $i$-th EV.



\subsubsection{Energy Balance}
EV charging sources can be classified into three categories: grid energy (G2V), V2V energy exchange, and locally sourced solar energy. Discharging from EVs can either be directed towards the grid (V2G) or towards other EVs.

The system under consideration includes multiple charging and discharging methods, such as V2G, V2V, and solar energy. During the charging process, each charger must determine both the direction and the amount of power. Hence, we introduce decision variables including the V2V charging power $p_{i,t}^\text{V2V,c}$, V2V discharging power $p_{i,t}^\text{V2V,d}$, power from the $i$-th EV to the grid $p_{i,t}^\text{V2G}$, power from the grid to the $i$-th EV $p_{i,t}^\text{G2V}$, power from photovoltaics (PV) to the grid $p_{t}^\text{PVG}$, and power from PV to the $i$-th EV $p_{i,t}^\text{PVEV}$. Initially, the charge/discharge balance is shown as follows\begin{align}
    & p_{i,t}^\text{ch} = p_{i,t}^\text{PVEV} + p_{i,t}^\text{V2V,c} + p_{i,t}^\text{G2V} ,\label{ch_b}\\
    & p_{i,t}^\text{disch} = p_{i,t}^\text{V2G} + p_{i,t}^\text{V2V,d}, \label{disch_b}\\
    & p_{i,t}^\text{PVEV}, p_{i,t}^\text{V2V,c}, p_{i,t}^\text{G2V},p_{i,t}^\text{V2G},p_{i,t}^\text{V2V,d}, p_{t}^\text{PVG} \geq 0. \label{no_zero}
\end{align}

The solar energy must satisfy the PV generation constraint
\begin{align}
    & 0 \leq  \sum_{i \in \mathcal{I}}p_{i,t}^\text{PVEV} + p_{t}^\text{PVG} \leq p_{t}^\text{PVgen}. \label{pvgen}
\end{align}

EVs purchasing and selling electricity from the grid must comply with the following constraints
\begin{align}
    & 0 \leq \sum_{i \in \mathcal{I}}p_{i,t}^\text{G2V} \leq G^\text{max},\label{g2v}\\
    & 0 \leq \sum_{i \in \mathcal{I}} p_{i,t}^\text{V2G} + p_{t}^\text{PVG}\leq G^\text{max}, \label{v2g}
\end{align}
where $G^\text{max}$ is the maximum energy transmission capacity between charging station and the grid.
Moreover, the V2V energy transfer must be balanced as
\begin{align}
    & \sum_{i \in \mathcal{I}} p_{i,t}^\text{V2V,c} = \sum_{i\in \mathcal{I}}p_{i,t}^\text{V2V,c}, \label{V2V_b}
\end{align}
in which the V2V energy transfer is divided to V2V power of consumer $p_{i,t}^\text{V2V,c}$ and V2V energy of producer $p_{i,t}^\text{V2V,d}$.

\subsubsection{User Satisfaction and Fairness Factor} 
In order to represent the satisfaction fairness model, this work applies the future average power (FAP) to represent user satisfaction, which means a constant power required by the electric vehicle to complete the charging task for the rest of the time. The EV charging satisfaction level is defined as follows
\begin{align}
    &FAP_{i,t} = \frac{E_i^\text{dem} - E_{i,t}}{T_{i,t}^r},\label{FAP}\\
    &U_{i,t} = -\rho \frac{FAP_{i,t}}{P^\text{ch,max}}, \label{sat}
\end{align}
where $\rho$ denotes a constant coefficient that balances the costs and completion of charging tasks. Then, the fairness factor $\psi_{i,t}$ can be defined as follows
\begin{align}
    &\psi_{i,t} = |U_{i,t} - \overline{U}_{t}|,\label{fair}
\end{align}
in which the fairness is represented as the distance between current EV charging satisfaction level $U_{i,t}$ and average satisfaction level $\overline{U}_{t}$ of all EVs.

\subsubsection{Objectives and Constraints}
According to the service of the charging station, this problem has three objectives: energy cost reduction, charging demand satisfaction, and fairness. The objective function can be expressed as
\begin{align}
\begin{split}
&\min ~~ \!\!\!\!\sum_{t\in \mathcal{T}}\sum_{i \in \mathcal{I}} \Bigl( (p_{i,t}^\text{G2V} \kappa_t^\text{buy} - p_{i,t}^\text{V2G}\kappa_t^\text{sell}) \Delta t \\
&+ AGE_{i,t}^\text{cyc} \kappa^\text{batt} + \psi_{i,t} - U_{i,t} \Bigr) - \sum_{t \in \mathcal{T}} (p_t^\text{PVG}\kappa_t^\text{sell} \Delta t)\\
&\text{s.t.}~~\eqref{xt_con} - \eqref{eit_con}~\text{and}~\eqref{ch_b} - \eqref{V2V_b},
\end{split}
\label{obj}
\end{align}
where $\kappa^{\text{buy}}_t$ and $\kappa^{\text{sell}}_t$ are energy selling price and energy purchasing price at time $t$. V2V energy exchanges cancel each other out in calculating the total cost.


\section{Multi-Agent Markov Decision Process}
This section formulates a Markov decision process (MDP) for the optimization problem in Section~\ref{sec:problem} and solves it using MARL. We consider each charger associated with an EV as an agent and demonte the set of $n$ chargers as $\mathcal{J}$, such that each individual charger index is given as $j \in \mathcal{J}$.

\textbf{State:} In the proposed model, the state of each agent is a data structure that includes components related to the calculation of costs/rewards, constraints, and any information necessary for the transition function. The states in our problem are divided into two categories: action-related states, denoted as $s_{j,t}^a$, and independent states, represented as $s_{j,t}^\text{ind}$. Action-related states, which are impacted by the agent's actions, corresponding to the present energy of the EV connected to the $j$-th charger, $E_{j,t}$. Moreover, independent states are environmental derivatives and encapsulate a variety of elements. These include the remaining time $T_{j,t}^r$, charging demand $E_{j,t}^{\text{dem}}$, solar energy supply $p_t^\text{PVgen}$, energy selling price $\kappa_{t}^\text{sell}$, and energy purchasing price $\kappa_{t}^\text{buy}$. The state of the $j$-th agent at time $t$ can be modeled as
\begin{align}
    &s_{j,t}^a {=} \{E_{j,t}\} \label{state_a},\\
    &\!s_{j,t}^\text{ind} {=} \{T_{j,t}^r, E_j^{\text{dem}}, \kappa_{t}^\text{buy},\kappa_t^\text{sell}, p_t^\text{PVgen}\}, \label{state_ind}\\
    &s_{j,t} {=} \{s_{j,t}^a, s_{j,t}^\text{ind}\}.  \label{state1}
\end{align}

\textbf{Action:} For the action of agents, each action should define the charging/discharging power and direction. The action of $j$-th agent at time $t$ can be expressed as
\begin{align}
    a_{j,t} = \{p_{j,t}, a_{j,t}^\text{V2V}, a_{j,t}^\text{PVEV}\}.
    \label{action}
\end{align}

The action space is continuous, the charging and discharging power must satisfy the constraints specified by the system model. To determine the source of energy, the $j$-th agent must transmit requests for V2V energy, $a_{j,t}^\text{V2V}$, and solar energy, $a_{j,t}^\text{PVEV}$, to the environment, illustrating its intention towards utilizing solar and V2V energy. Based on the proportion of energy capacity and requested energy for solar and V2V energy, the environment will allocate the actual solar power, $p_{j,t}^\text{PVEV}$, and the V2V energy transfer, $p_{j,t}^\text{V2V,c}$/$p_{j,t}^\text{V2V,d}$, to each EV, which follows the constraints of the system. When $p_{j,t}^\text{PVEV}$ and $p_{j,t}^\text{V2V,c}$/$p_{j,t}^\text{V2V,d}$ are determined, the V2G power $p_{j,t}^\text{V2G}$, G2V power $p_{j,t}^\text{G2V}$ and PV to grid power can be calculated based on \eqref{ch_b}-\eqref{pvgen}.

\textbf{State Transition:} The state transition is a set of rules for the state of agents after receiving an action. In this model, we set all state elements as 0, when the charger is empty. If the EV charger is working, the state transition can be expressed as
    \begin{align}
            &T_{j,t+1}^r {=} T_{j,t}^r - \Delta t, \label{RT}
    \end{align}
where $T_{j,t+1}^r$ is remaining time of next time step, and $E_{j,t+1}$ can be calculated by \eqref{eit}. The next state information $\kappa_{t+1}^\text{buy}$, $\kappa_{t+1}^\text{sell}$, $E_j^{\text{dem}}$ and $p_{t+1}^\text{PVgen}$ can be updated by real-time data.

\textbf{Reward:} Reinforcement learning aims to develop an optimal strategy by maximizing the total reward, which is also a crucial aspect of artificial general intelligence \cite{silver2021reward}. The design of the reward function is intrinsically connected to the objectives of the problem at hand. In this study, the goal is to minimize the total cost and meet the charging demands of the EVs upon their departure from the station.

As such, the reward function must account for three key components: energy cost, user satisfaction and the fairness metric.
\begin{itemize}
\item \textbf{Cost: }
The overall cost includes both energy cost and battery degradation cost. Agents need to cooperate to achieve the goal of reducing the total cost. As such, the reward structure is designed to assign the average cost to each agent, which can encourage all agents to work collectively towards cost reduction with fluctuating solar and energy prices. According to the objective function \eqref{obj}, the reward for the $j$-th agent in terms of the total cost, included the EV energy cost $R_{t}^\text{energy}$, the benefit from selling solar energy to the grid $R_{t}^\text{PVG}$ and battery degradation cost $R_{t}^\text{battery}$ at time step $t$. The average cost $R_{j,t}^\text{cost}$ for the $j$-th agent at $t$ can be represented as
\begin{align}
    & R_{j,t}^\text{cost} {=} - (\frac{R_{t}^\text{energy} - R_{t}^\text{PVG} + R_{t}^\text{battery} }{n_t}).
\end{align}
In this context, the positive cost equates to the negative reward within the RL model. Additionally, the quantity of active agents at time $t$ is represented as $n_t$.

\item \textbf{User Satisfaction and Fairness: }
For EV user experience, the reward $R_{j,t}^\text{user}$ can be calculated via charging satisfaction level \eqref{sat} and fairness \eqref{fair} as
\begin{align}
    & R_{j,t}^\text{user} = U_{j,t} + \psi_{j,t}.
\end{align}
\end{itemize}
At last, the MARL needs to deal with the multi-objective problem, which is to minimize the total charging cost (with battery degradation cost) while satisfying the charging demand considering fairness. Thus, we set weighted reward in the total reward calculation. Thus, the total reward can be calculated as 
\begin{align}
    & R_{j,t} = \xi R_{j,t}^\text{cost}  + (1-\xi)R_{j,t}^\text{user} + R^\text{grid},
\end{align}
where $\xi$ is the trade-off parameter between charging demand and costs. The penalty of grid constraint violation is $R^\text{grid}$.

\section{Proposed MARL Method}
In this section, we present the MARL algorithm, which orchestrates EV charging to optimize the cumulative system reward. Effective information exchange among chargers is needed during the training process. We utilize the Multi-Agent Deep Deterministic Policy Gradient (MADDPG) algorithm with Centralized Training with Decentralized Execution (CTDE) scheme \cite{lowe2017multi}, in which the policy is trained centrally and actions are independently executed for each agent in a decentralized manner. The adopted framework is enabled by the Actor-Critic (A2C) architecture. The actor generates the EV charging action via policy neural network, while the critic evaluates the respective actor's actions via critic neural network.

\begin{figure}[t]
    \centering
    \includegraphics[width=8.5cm]{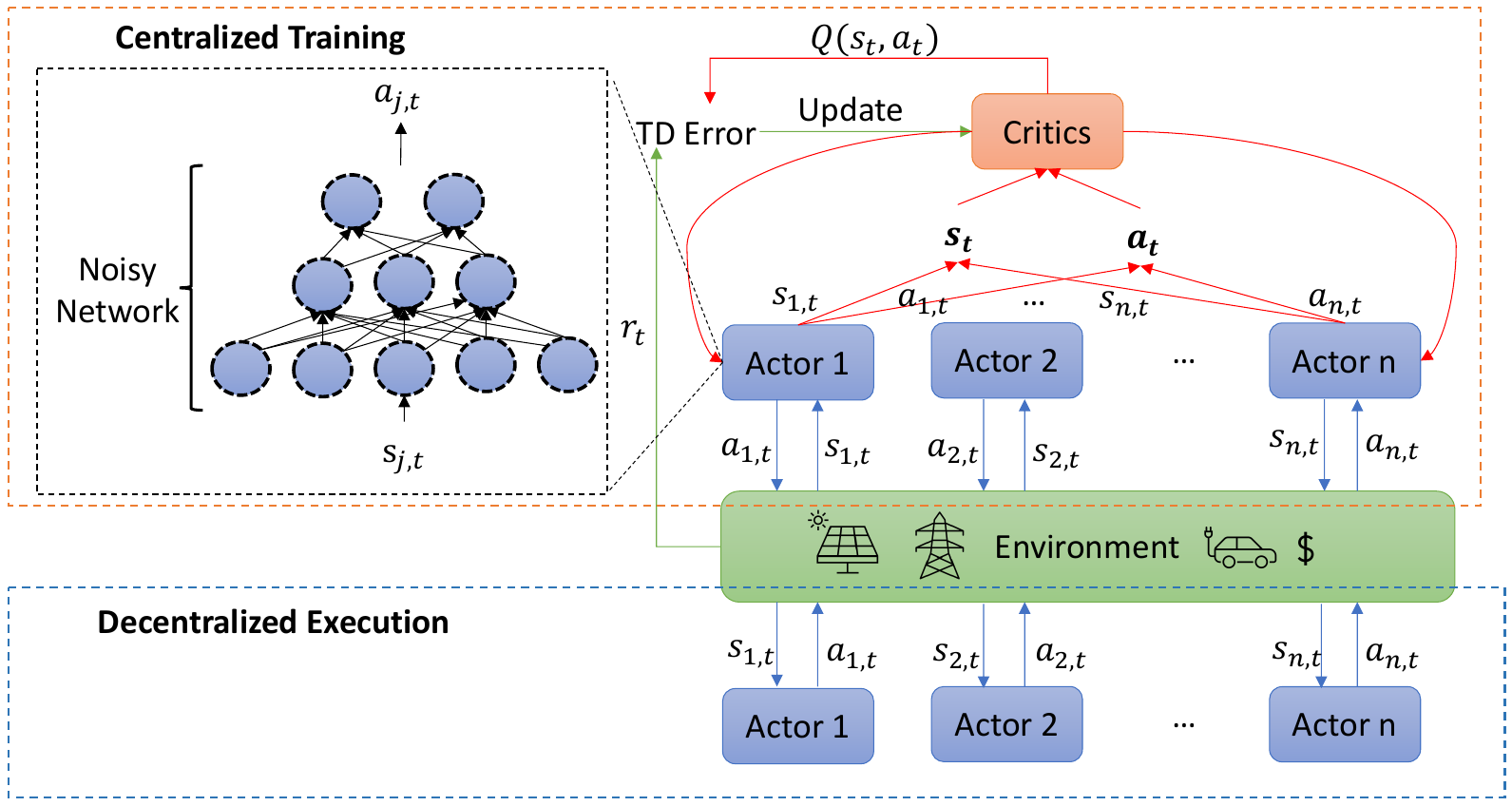}
    \caption{The architecture of CTDE with noisy network.}
    \label{algorithm}
\end{figure}

The structure of CTDE architecture is demonstrated in the right-hand section of Fig. \ref{algorithm}. The training process remains centralized, which must access the information from other chargers. The centralized  critic utilizes the critic neural network to approximate the Q-function $Q_j^\mu(\boldsymbol s, a_1,...,a_n)$ for policy $\mu$. Conversely, actors are restricted to local information only, using actor neural networks $\mu_j$, parameterized by $\theta_j$ to produce action under current state $s_j$. 
In centralized training, Temporal Difference (TD) learning is used to estimate the centralized 
Q-function and construct TD-target $y$. According to \cite{lowe2017multi}, the loss function of the critic (TD-error) can be expressed as
\begin{align}
    &\mathscr L(\theta_j) {=} \mathbb{E}_{s,a,r,s'}[(Q_j^\mu(\boldsymbol s, a_1,...,a_n) - y)^2] \label{td_error},\\
    &y {=} r_j + \gamma Q_j^{\mu'}(\boldsymbol s', a_1',...,a_n')\big|_{a'_k=\mu'_k(s_k)} \label{td_target},
\end{align}
where the target policy network is denoted as $\mu_j'$ with delayed parameter $\theta_j'$ and next state is represented as $\boldsymbol s'$.  The TD target $y$ is an estimation of the cumulated reward for the next step based on the most recent reward. The loss function, also known as the TD-error, quantifies the difference between the current estimation and the estimation from the previous step.

When considering the Actor (policy) networks, we aim to learn a policy that operates under a continuous action space, which can be achieved through the deterministic policy gradient. The loss function can be formally articulated as
\begin{align}
    \begin{split}
    &\nabla_{\theta_j}J(\boldsymbol \mu_j) {=} \mathbb{E}_{s\sim p,a_j\sim D}\\&[\nabla_{\theta_j} \boldsymbol \mu_j(a_j|s_j)\nabla_{a_j}Q_j^{\boldsymbol \mu}(\boldsymbol s|a_1,...,a_n)  \big|_{a_j=\boldsymbol \mu_j(s_j)}],
    \end{split}
\end{align}
where $D$ is the experience replay buffer that contains the tuples $(s,s', a_1,...,a_n, r_1,...,r_n)$ for all agents.

Also, the parameter noise is added to the neural network for better exploration. We modify the original algorithm by incorporating a noisy network \cite{plappert2017parameter}. This addition involves substituting the standard linear layer with a noisy linear layer, which is described as follows
\begin{align}
    & Y = (\nu^\theta + \sigma^\theta \odot \epsilon^\theta)X + (\nu^b + \sigma^b \odot \epsilon^b),
\end{align}
in which $\epsilon = [\epsilon^\theta, \epsilon^b]$ represents randomly sampled noise matrices exhibiting zero mean and fixed statistics. In addition, $\nu = [\nu^\theta, \nu^b]$ and $\sigma = [\sigma^\theta, \sigma^b]$ denote the neural network's learnable parameters for weights and bias. Within the noisy network framework, instead of applying an $\epsilon$-greedy policy, the agent is capable of acting greedily in accordance with a network utilizing noisy linear layers.

Upon completion of the training phase, the policy networks are able to approximate the optimal policy function. Given that the policy network is embedded within the decentralized agent, it enables the agent to make autonomous decisions during the execution process without dependency on other agents' information.

\section{PERFORMANCE EVALUATION} \label{sec:eval}
To simulate EV charging processes, we use EV data from the Adaptive Charging Network (ACN) dataset \cite{lee2019acn} with a fixed charging/discharging power limits set at $16$kW. We employ corresponding solar data to simulate the energy generation of our solar power system, whose solar capacity is configured at $30$kWp. In order to compute the energy cost for the charging station, EVs purchase energy from grid following a Time-of-Use (TOU) pricing scheme and sell energy to grid using the wholesale energy price. We use real price data from National Electricity Market (NEM) in Australia \cite{aemo_dashboard}. We assume that the usage of solar energy is free for all EVs in the charging station. According to a previous study \cite{shurrab2021efficient}, the V2V price should ideally be bracketed between the selling price for the producer and the purchased price for the consumer, facilitating the participation of EVs in V2V energy exchange activities. In this study, the V2V price is assigned as the midpoint of the TOU energy price and the wholesale market energy price.

In the experiments, we consider $20$ chargers (i.e., agents) and design a centralized model-based optimization algorithm as a baseline. This baseline method is Rolling-Horizon Optimization (RHO), which relies the predictions of solar energy generation and energy prices to determine the EV charging and discharge schedule in a look-ahead window (as the optimization horizon) and execute the decision in the current time slot. The window size of the rolling horizon aligns with the longest EV parking time upon the arrival of a new EV. Morevoer, in contrast to the model-based RHO baseline, we also design a model-free baseline, which is a centralized multi-agent deep Q-network (MADQN) for comparisons with our proposed MADDPG with CTDE.

\begin{figure}[t]
    \centering
    \includegraphics[width=6.5cm]{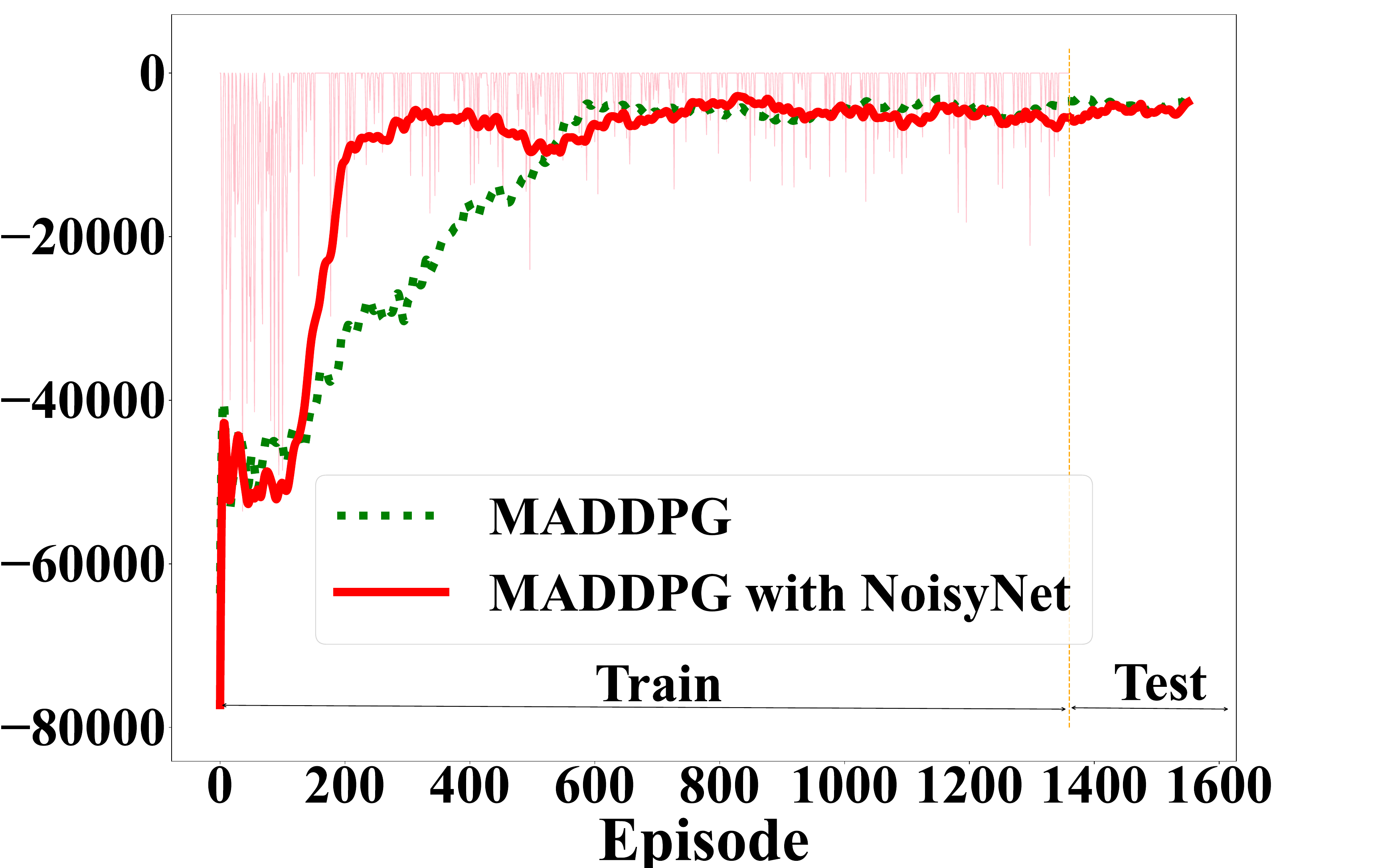}
    \caption{Effect of Noisy Network on Algorithm Convergence.}
    \label{conv_noise}
\end{figure}

Fig. \ref{conv_noise} provides a comparative depiction of the convergence of the MADDPG with and without incorporating a noisy network-NoisyNet. We see that MADDPG with NoisyNet converges faster. The result suggests that incorporating NoisyNet can enhance the exploration capabilities of agents within the environment. Hence, the agents become proficient in learning an optimal policy within a limited number of episodes.

\begin{figure}[t]
    \centering
    \includegraphics[width=8.8cm]{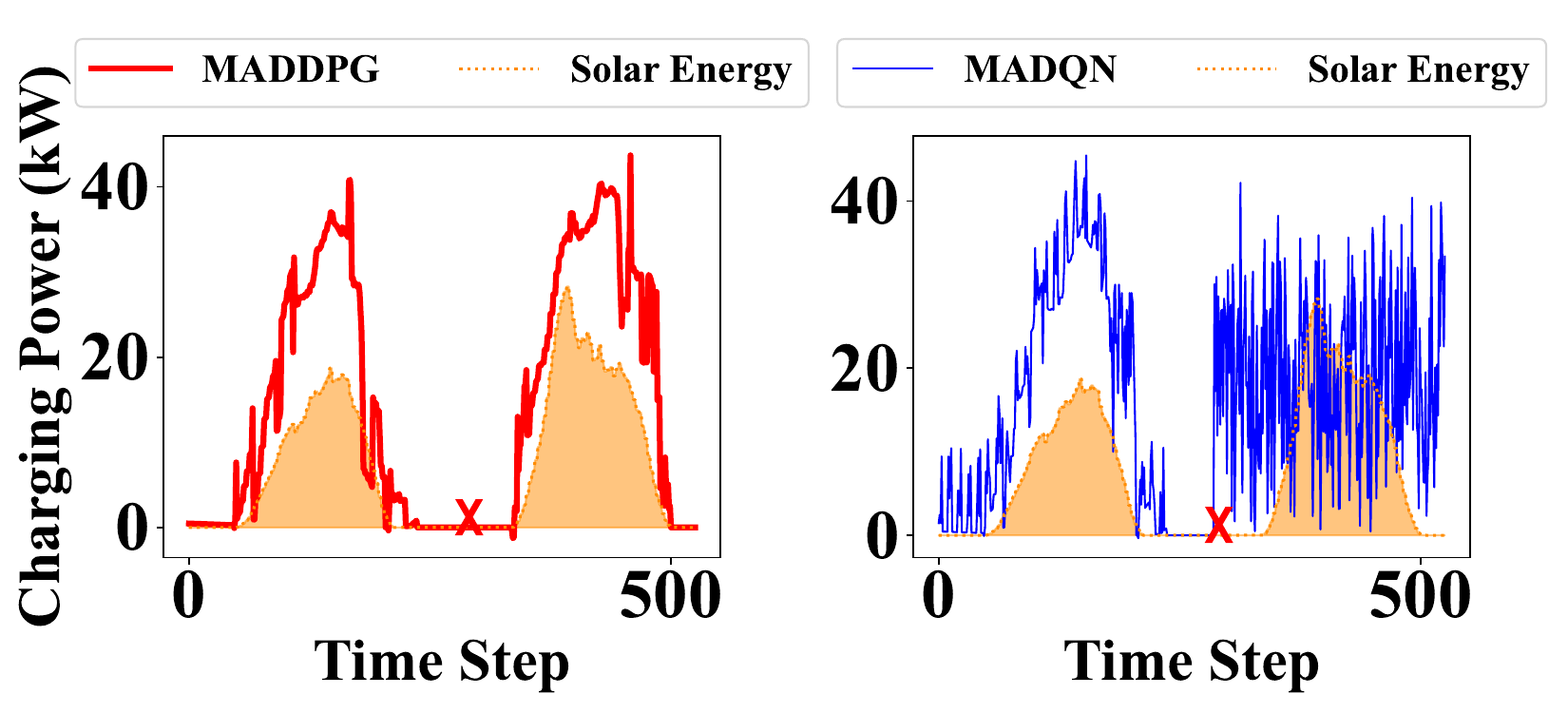}
    \caption{The performance between centralized and decentralized MARL in the case of partial faults.}
    \label{Den}
\end{figure}

The effectiveness of the CTDE framework is demonstrated in Fig. \ref{Den}. The dotted orange curve in the figure exhibits the real-time solar energy generation. We see that the MADDPG enables the agent to learn the pattern of solar energy generation and utilize solar energy to reduce energy cost. This is evidenced by the alignment of the EV charging (in red solid curve) and the available solar energy. Furthermore, the red cross symbolizes the occurrence of a partial system fault, simulating failures in some chargers by replacing the information of those faulty chargers with random numbers. A partial system fault event is introduced after the 280th time step. Upon comparing the centralized and decentralized charging power, it becomes evident that the partial fault leads to instability of the remaining chargers using the centralized MADQN, whereas the MADDPG continues to function normally.

\begin{figure}[t]
    \centering
    \includegraphics[width=6.5cm]{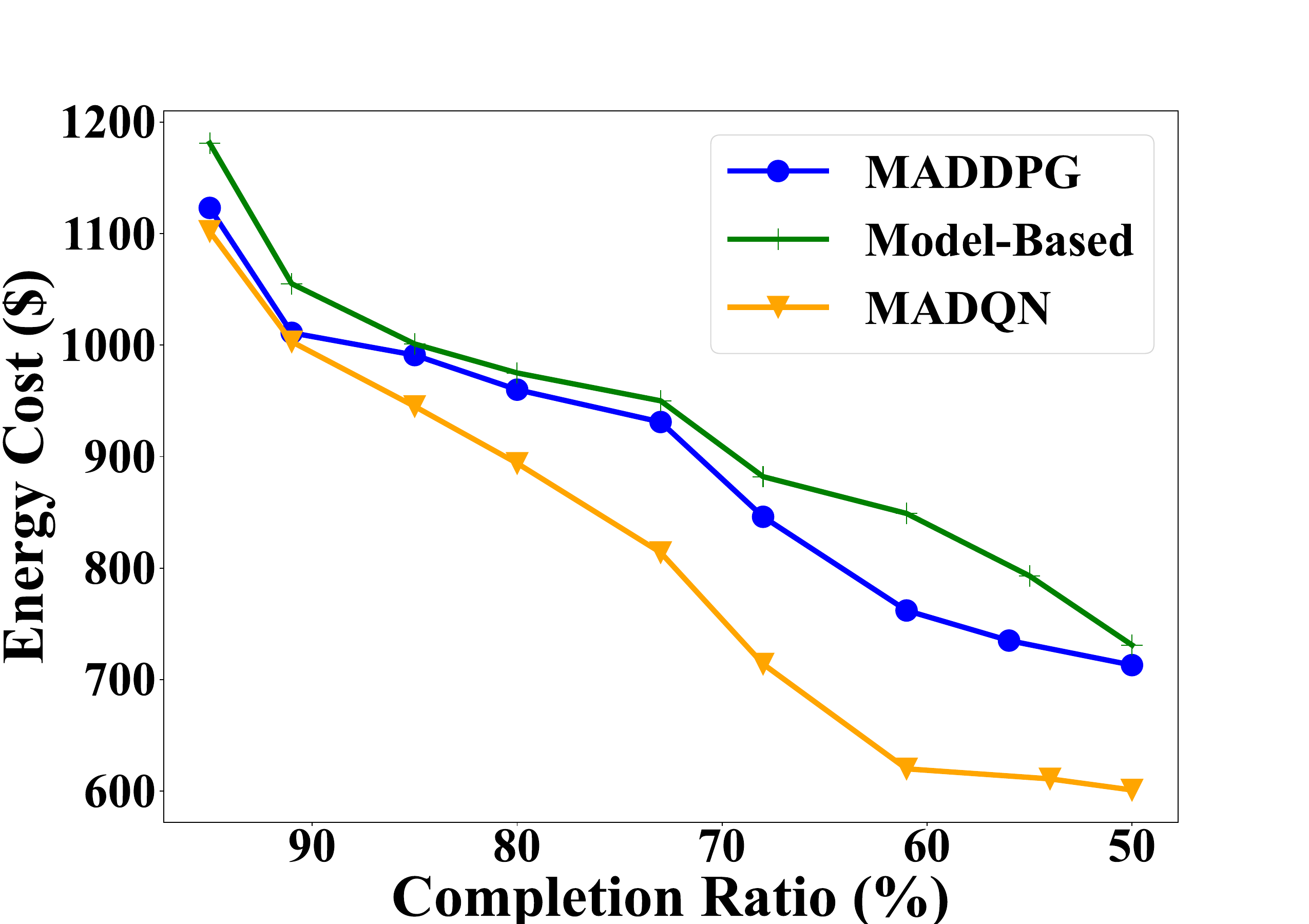}
    \caption{Tradeoff between energy cost and user satisfaction.}
    \label{v2v_cost}
\end{figure}

Since our system includes multiple objectives that the charging station and EV owners care about, such as energy cost, user satisfaction, and fairness, it becomes necessary to establish trade-offs in these distinct objectives. In our work, we use the charging task completion ratio to represent user satisfaction, which is the percentage of the completion across all charging tasks. Fig. \ref{v2v_cost} depicts the trade-off between the user satisfaction and energy cost. In Fig. \ref{v2v_cost}, a lower position indicates better performance. The results show that the MARL algorithms outperform the model-based approach. Comparatively, MADDPG outperforms the model-based approach while exhibiting slightly compromised performance than the MADQN.

\begin{figure}[t]
    \centering
    \includegraphics[width=6.5cm]{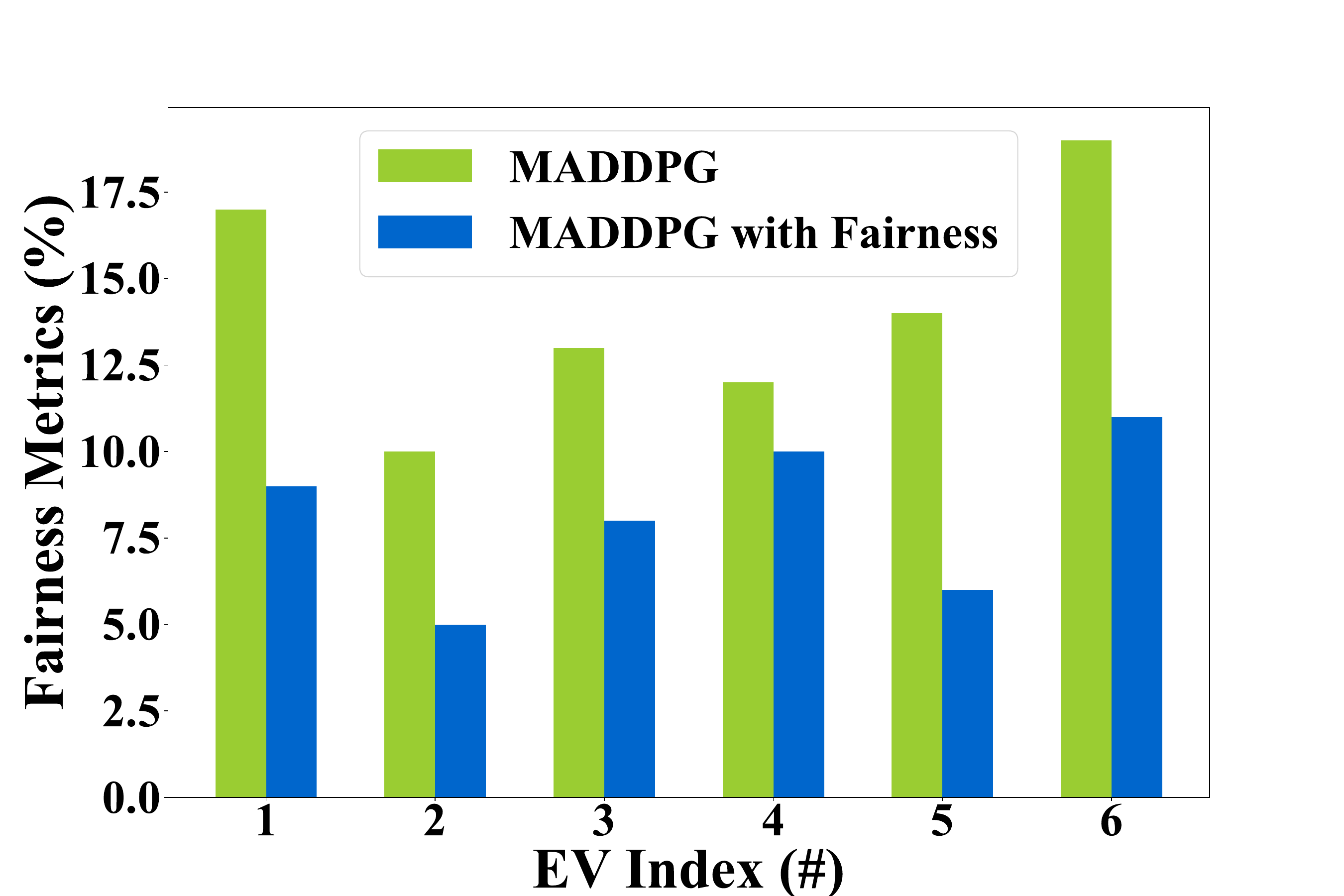}
    \caption{Fairness of charging strategies.}
    \label{fair_g}
\end{figure}

Moreover, our study incorporates considerations of fairness with respect to the user satisfaction. A fairness factor is integrated into the MARL reward mechanism. In order to quantify the level of fairness, we compute fairness metrics for a chosen set of EVs, which is the distance between the completion ratio of each selected EV and the average completion ratio. According to Fig. \ref{fair_g}, the result suggests that the MADDPG with fairness maintains lower metrics, signifying a relatively smaller disparity in user satisfaction across different EVs.

The scalability of the MADDPG with NoisyNet is examined under varying scenarios, encompassing five different sizes of the charging station. These results are presented in TABLE \ref{scala}. With an increase in the size of the charging station (i.e., the number of chargers), the average energy cost per each charger exhibits relative consistency. Similarly, the average completion ratio of each charger demonstrates remains at a similar level.
\begin{table}[t]
\caption{Performance of MADDPG under different station size}

\setlength{\tabcolsep}{1.8mm}{\begin{tabular}{|c|c|c|c|c|c|}\hline
Number of chargers &10&20&40&60&80\\\hline
Average Completion Ratio &81&84&80&84&82\\\hline
Average Energy Cost &13.23&11.05&11.29&10.51&10.08\\\hline
\end{tabular}}
\label{scala}
\end{table}

\section{Conclusion}
This paper proposed a decentralized MARL approach to coordinate EV charging with V2V and solar energy integration. The proposed algorithm seeks to minimize the total energy costs while enhancing the EV user experience in terms of charging satisfaction and fairness. Leveraging the capability of MARL to learn from EV charging environment, EV chargers, modeled as agents, make independent decisions without information exchange. This unique attribute ensures operational normalcy during instances of partial system faults. Additionally, the inclusion of a noisy network within the MARL algorithm fosters agent exploration of the environment, leading to faster convergence. Numerical studies demonstrated that the proposed algorithm exhibited scalability and superior performance compared to traditional optimization baselines. For future research, we will consider the integration of EV charging station into the grid to provide ancillary services with network constraints.


\bibliographystyle{IEEEtran}
\bibliography{IEEEabrv.bib,ref.bib}

\end{document}